# Deciphering core-exciton dynamics in CaF$_2$ with attosecond spectroscopy

Rafael Quintero-Bermudez and Stephen R. Leone

*Department of Chemistry, University of California, Berkeley, California 94720, USA*
*Chemical Sciences Division, Lawrence Berkeley National Laboratory, Berkeley, California 94720, USA*
*Department of Physics, University of California, Berkeley, California 94720, USA*

Core excitons in solids have garnered increasing interest, yet their behavior and decay mechanisms are not fully understood. Here, we use attosecond extreme ultraviolet (XUV) transient absorption spectroscopy, performed with a broadband 25-45 eV sub-fs XUV pump pulse and a 500-1000 nm sub 5 fs near-infrared (NIR) supercontinuum probe pulse to monitor the excitation, dynamics, and decay of core excitons in CaF$_2$ at the Ca$^{2+}$ M$_{2,3}$ edge. The XUV pulses are used to excite core excitons in CaF$_2$ based around the Ca$^{2+}$ and the polarization of the medium is subsequently perturbed by the time-delayed NIR pulses to measure the spectral changes and decays. A number of features are identified in the transient absorption spectrum, which suggest transfer between excitonic states, Stark shifts, and the emergence of light-induced states. We find that various core excitons identified exhibit coherence lifetimes spanning 3-7 fs. Furthermore, a NIR-intensity-dependent analysis finds a negative correlation with the coherence lifetime of various identified excitonic features, supporting a phonon-mediated mechanism as responsible for the core exciton decoherence. We present a computational band structure projection analysis strategy to estimate the orbital structure of the core excitons and determine which core excitonic transitions should be allowed by selection rules with the probe beam. This strategy is found to successfully describe the observed spectroscopic data. The outlined joint spectroscopic and computational investigation of core excitons is a powerful technique that explains the complex behavior of core excitons in solid-state materials.

## I. INTRODUCTION

Excitons, condensed matter quasi-particles that consist of a bound electron-hole pair [1], have been studied since the early seminal work of Frenkel [2], Wannier [3], and Peierls [4]. Exciton research has led to remarkable achievements in photovoltaics [5], light-emission [6], solar water-splitting [7], and a wide range of electronic devices [8]. Excitons that consist of a bound electron and a core hole, denominated core excitons, were first identified in extreme-ultraviolet (XUV) and x-ray absorption spectra in the early 1970s due to increased accessibility of synchrotron radiation sources [9,10]. However, detailed understanding of the short-lived dynamics and decoherence of core excitons has only recently become possible with the development of few-femtosecond and attosecond XUV pump-probe spectroscopy techniques with tabletop high harmonic generation (HHG) sources [11–15]. HHG sources produce broad-spectrum XUV and soft x-ray pulses capable of supporting attosecond temporal resolution [16].

Core excitons are best observed in ionic insulators due to their relatively low dielectric constant and consequently weak screening effects. Core excitons have been observed in a wide range of large bandgap insulators [9]. Their absorption is often characterized by distinct strong and narrow absorption lines clearly visible at room temperature at a considerably large energy below the conduction band absorption, a quantity known as the binding energy. CaF$_2$ has been known to display a particularly distinct and strong absorption line in its XUV absorption spectrum [9,17,18], but the excitonic nature of this feature could not be confirmed. Furthermore, additional neighboring features were identified that also could not be conclusively identified as excitonic absorption.

Ultrafast pump probe spectroscopy studies have been conducted to investigate core excitons in only a handful of materials [11–14]. CaF$_2$ is an intriguing candidate to study core excitons given the doubly charged nature of the cation, its amenability to calculation, its rich XUV absorption spectrum and the lack of data in the literature on its interpretation. Additionally, neither the Ca core levels nor fluorite crystal structure core excitons have been investigated, further motivating this study. The CaF$_2$ crystal structure is shown in Fig. 1(a): inset.

In many typical pump-probe XUV spectroscopy schemes, a near-infrared (NIR) or visible pump pulse is used to excite carriers in a sample, such as electrons across a bandgap in semiconductors and insulators. This is followed by an XUV probe pulse to monitor time-resolved changes in the absorption spectrum. Core excitons, however, necessitate XUV or x-ray pump photons and the reversal of this pulse scheme. As a result, an attosecond XUV pulse is used first to pump core excitons and the NIR pulse is used second as the probe by perturbing the core exciton transition dipoles or exciting to neighboring states. These measurements are not at risk of becoming overshadowed by valence band (VB) to conduction band (CB) carrier excitation due to the very large-bandgap in insulators such as CaF$_2$.

Here, we report on the core exciton dynamics at the Ca$^{2+}$ M$_{2,3}$ edge in the alkali-halide insulator CaF$_2$. This study consists of ultrafast XUV transient absorption spectroscopy (UXTAS) utilizing the XUV pump and NIR probe. The transient absorption measurements depict a rich array of time-dependent features in the 27-37 eV range. The features are used to identify resonant absorption of core-excitonic states, formation of light-induced states, Stark shifts, and charge transfer between core excitonic states. Evidence is presented for the decay of the core excitonic features to be interpreted as phonon-mediated. A computational band structure projection strategy, consisting of density functional theory (DFT) and Bethe-Salpeter-equation (BSE) analysis, is presented to estimate the orbital structure of core excitons and to determine which core excitonic transitions would be dipole allowed by selection rules with the NIR probe beam. The outlined joint UXTAS and computational band structure projection analysis investigation of core excitons is a powerful technique that satisfactorily explains the complex behavior of core excitons in solid-state materials.

## II. EXPERIMENTAL SCHEME

A detailed account of the experimental details is included in Appendix A. In brief, a sub-5 fs 500-1000 nm NIR probe pulse is incident on samples with a time delay after a 25-45 eV XUV probe pulse that is generated by HHG using the aforementioned NIR pulse incident onto a Kr gas cell. This NIR pulse modifies the decay of the polarization of the medium set up by the initiating XUV pulse. The time delay between the pulses is controlled by means of a retroreflector mounted onto a delay stage. The XUV beam is spectrally dispersed by a grating and recorded with an X-ray camera. Samples consist of 25 nm thick $CaF_2$ thin films evaporated onto 30 nm thick $Si_3N_4$ membrane substrates. The samples were characterized by powder X-ray diffraction to confirm the crystalline structure of the sample [Fig. S3].

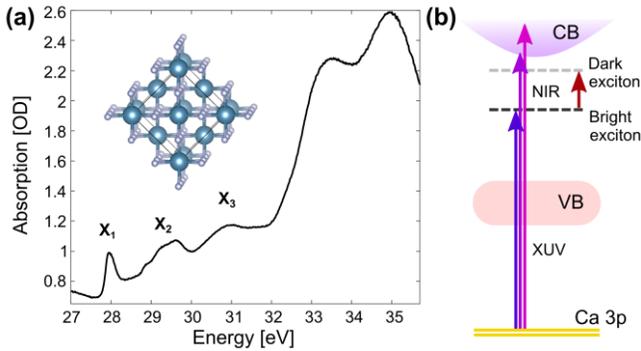

FIG 1. Linear extreme ultraviolet (XUV) absorption spectrum of $CaF_2$ at the Ca $M_{2,3}$ edge. a) Linear absorption spectrum of a $CaF_2$ thin film (25 nm) used in the transient absorption study. Inset: Crystal structure of $CaF_2$. b) Schematic depiction of energy levels and bands involved in XUV pump - NIR probe experiments.

## III. RESULTS

### A. Transient Absorption Spectroscopy

The XUV absorption spectrum for a 25 nm thick $CaF_2$ thin film is shown in Fig. 1(a). The distinct absorption peak at 28 eV (labelled $X_1$) has been attributed to a $\Gamma$-point core exciton forming between the photoexcited electron and the $Ca^{2+}$ 3p core hole [9]. Hereafter, $Ca^{2+}$ is simply indicated as Ca. The Ca 3p core levels are spin-orbit split but their energy separation is negligible such that the absorption spectrum cannot distinguish them [9]. Between 28.6-30 eV and 30-32 eV (labelled $X_2$ and $X_3$, respectively), a series of features are clearly observed. In previous work, these features were barely noticeable and were not interpreted. Here, we posit that these are also core excitons consisting of electrons from higher energy CB levels in $CaF_2$, and verify this assumption below via the transient spectra. The baseline absorption can be seen to increase after the $X_1$ feature. This is likely due to non-excitonic excitation from the Ca 3p core levels to the CB. Finally, absorption is seen to increase significantly above 32 eV. This can be attributed to an increase in CB density as well as the excitation of F 2s core excitons to the CB.

A schematic of the UXTAS measurement is shown in Fig. 1(b). A broadband XUV pulse pumps Ca 3p core electrons into the CB and core excitons. The excited core excitons are subsequently probed with the NIR probe pulse. As mentioned earlier, the bandgap of $CaF_2$ is far too large for electrons to be excited from the VB to the CB with the NIR pulse.

The UXTAS spectrum of the $CaF_2$ thin film collected with the highest NIR beam intensity or pulse energy used (10.2 TW/cm$^2$ or 12 µJ, respectively) is shown in Fig. 2(a). As outlined above, positive time means that the XUV pulse arrives before the NIR pulse. Distinct transient features can be observed across the entire 27-37 eV range. Despite the high NIR intensity used, the observed transient absorption features are comparably weaker than those observed in previously studied core excitonic systems, such as MgO [12]. Most features only seem to appear for the duration of the overlap of the XUV with the 5 fs NIR pulse. Three longer-lived features are observed that match the locations of $X_1$, and two regions within $X_2$. These are labelled A, B, and C in Fig 2(a), respectively. In order to facilitate analysis, snapshots of the transient absorption at various time delays are presented as the change in absorption (or $\Delta OD$) [Fig 2(b)] and as total absorption [Fig. 2(c)]. The latter is presented with the $\Delta OD$ component multiplied by a factor of 10 and adding this to the static absorption to facilitate the observation of the weak changes in absorption.

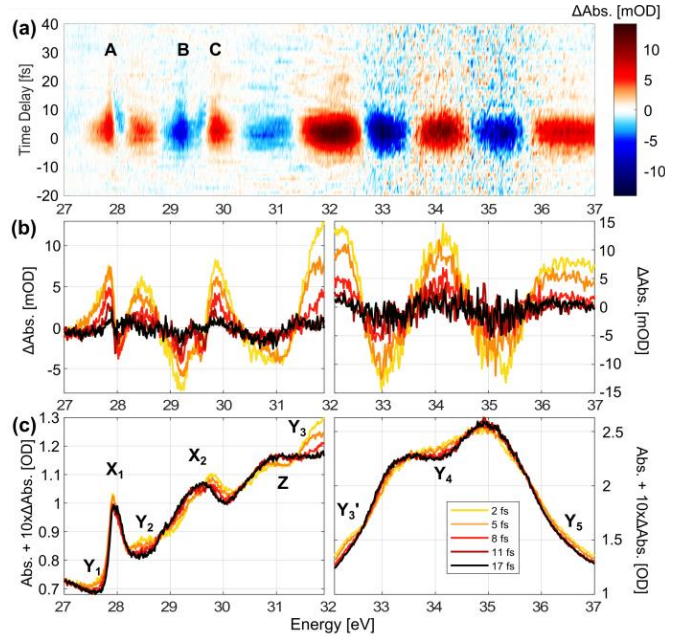

FIG 2. Few-femtosecond dynamics of core excitons in $CaF_2$ observed by XUV transient absorption spectroscopy. a) Delay-dependent transient absorption. Labels A, B, and C, highlight longer-lived features, identifying them as core excitons. b) Lineouts of delay-dependent transient absorption and c) total absorption spectral traces taken at 2, 5, 8, 11, and 17 fs. The latter is plotted by adding the transient absorption component magnified by a factor of 10 to the static spectrum to facilitate observation of weak features. Labels $X_n$, $Y_n$, and Z identify excitonic features, photoinduced absorption, and bleaching, respectively.

Three types of features are identified: (1) increases in absorption at features $Y_1$, $Y_2$, $Y_3$, $Y_4$, and $Y_5$, (2) a bleaching of excitonic feature $X_3$, labelled Z, and (3) shifts in excitonic absorption features $X_1$ and $X_2$. The $Y_n$ photoinduced absorption features are ascribed to the emergence of light-induced states, as has been done for similar features in previous core exciton dynamics studies on MgO [12]. $Y_{3-5}$, which consist of features well into the CB, however, could also result from renormalization of the bandgap upon photoexcitation [19]. The bleach at Z might result from charge transfer from nearby core excitons, such as $X_1$ or those within the $X_2$ set of core excitons. There is an additional feature at 29.2 eV that could be attributed in part as a bleaching resulting from charge transfer. However, it is difficult to distinguish this potential effect from the shifts that are also observed. To determine the source of the excitonic shifts, the same experiment was conducted at various NIR intensities. In order to ascertain the shift in energy, features $X_1$ and $X_2$ are fitted by a sum of Gaussian functions in order to determine the shifts incurred by each component.

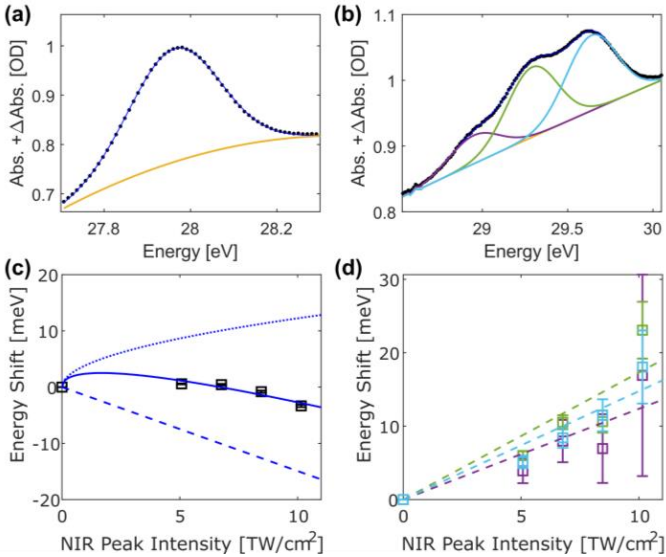

FIG 3. Fitting of core excitonic features at various NIR peak intensities reveal nature of spectral shifts. Fitting to one and three Gaussian functions, respectively, is used to fit total transient absorption spectra in excitonic features a) $X_1$ and b) $X_2$ after a subtraction of the baseline absorption. c) and d) Energy shifts as a function of NIR peak intensities obtained from Gaussian fitting of excitonic features a) $X_1$ and b) $X_2$, respectively. Fitting consists of a positive sublinear term (excitonic transfer: dotted line) and negative linear term (Stark: segmented line) (total: full blue line) in $X_1$ and a sole linear term is found to describe energy shifts in $X_2$ adequately (Stark: segmented lines, colors in b)).

The results of the Gaussian fitting are presented in Fig. 3(a) and (b), for features $X_1$ and $X_2$, respectively. The energy shifts of each Gaussian component are plotted as a function of NIR peak intensity in Fig. 3(c) and (d), for features $X_1$ and $X_2$, respectively. The core exciton shifts could result from the optical Stark effect [20] or could instead indicate transfer between core excitonic states due to the NIR pulse [12]. The former is known to exhibit a linear increase with NIR intensity, whereas the latter is expected to exhibit a sublinear increase with NIR intensity, as noted in [12].

The measured peak shifts in feature $X_1$ are best described by a sum of a positive sublinear and negative linear shift, which would correspond to a redshift due to the optical Stark effect and a blueshift due to excitonic transfer by the NIR field. From these observations, it can be inferred that the NIR energy (of 1.2-2.5 eV) is smaller than the energy of a dark exciton D but that some transfer (albeit a small proportion of the excitonic population of $X_1$) is able to take place between these two features. The measured peak shifts in feature $X_2$ are all sufficiently well fitted by positive linear increases, i.e., blueshift due to an optical Stark effect. In this case it is inferred that the NIR energy is smaller than the energy of a dark exciton D' and that transfer between $X_2$ core excitons and D' must be negligible. This experiment, supported by the bleach of feature $X_3$ [labelled Z in Fig. 2(c)], suggests the existence of a dark exciton at approximately 31 eV.

### B. Core Exciton Decay

The decoherence of core excitons in insulators has been debated in the literature since their discovery. Auger decay [21], electron-phonon coupling [12], electron-electron interactions [14], and mixing of core exciton states [14], are a few of the most promising causes discussed in the literature. In order to provide evidence that might support a mechanism for core exciton decoherence, the decay lifetimes of the three longer-lived features, albeit all three extremely short-lived, in the UXTAS [A, B, and C in Fig. 2 (a)] were determined for various NIR intensities. Singular Value Decomposition was used around features A, B, and C to ascertain the time-dependent trace. The decay of these features was then fitted with an exponentially modified Gaussian function in time to account for the instrument response function, which exhibits a Gaussian lineshape due to the shape of the NIR pulse.

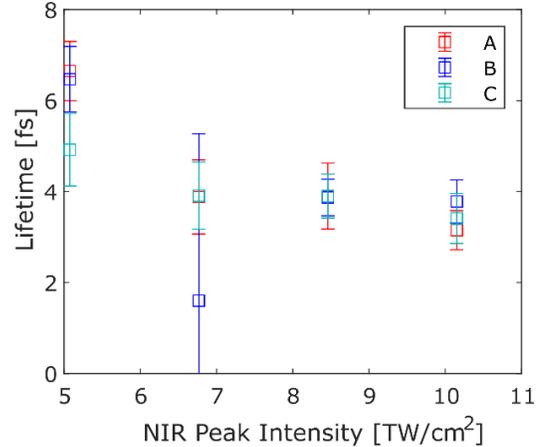

FIG 4. Coherence lifetimes of core excitonic features show a negative correlation with NIR peak intensity supporting a phonon-mediated decay. Lifetimes obtained from fitting the main transient component with a Singular Value Decomposition of data around features A, B, and C to an exponentially modified Gaussian distribution function.

The measured coherence lifetimes of the core excitons indicate a negative correlation with the NIR peak intensity. Although this property cannot single-handedly confirm a specific decoherence mechanism, it does support a phonon-mediated decay mechanism [11,12,22,23]. This is further discussed in the Discussion section IV.B below.

### C. Computational modelling

DFT computations were used to understand the UXTAS results above. In particular, DFT was used to determine the conduction band levels that could be associated with each core excitonic feature and the allowed transitions between them.

Firstly, DFT computations were completed with the Quantum Espresso software package [24,25] and were used to determine the projected band structure of $CaF_2$. This computation is illustrated in Fig. 5 depicting the projection of 5 core levels (Ca 3p and F 2s), all 6 VB levels, and 6 CB levels onto 6 electron orbitals: Ca 3s, Ca 3p, Ca 4s, Ca 3d, F 2s, and F 2p. It is observed that the core levels consist primarily of their expected electron orbitals, whereas the VB consists primarily of F 2p character. The CB is almost entirely Ca 3d type, with the exception of the CB edge near the Γ-point which corresponds to Ca 4s type and higher up CB levels (particularly in the vicinity of the Γ- and L-points) which exhibit some F 2p character.

Second, DFT computations with BSE were conducted with the Exciting computational package [26] to determine the excitonic weight contributions of band structure levels to the linear absorption excitonic peaks. The result of these computations is illustrated in Fig. 6. Here, the computed linear absorption spectrum is shown in Fig. 6(a), which closely resembles the measured XUV absorption spectrum. On this figure, 8 labels are placed throughout the spectrum highlighting representative spots within core excitonic features. Next, wavefunctions are illustrated in real space for the first two highlighted spots within $X_1$ in the calculated absorption spectrum, Fig. 6(b) and (c). Interestingly, the shapes of these wavefunctions resemble s and d orbitals, respectively. It is found, however that excitonic weights plotted onto the band structure are more informative as to the orbital shape of the excitons, especially when coupled with the projected band structure in Fig. 5. With this in mind, band structures depicting excitonic weights are shown for all 8 highlighted points in the XUV absorption spectrum in Fig. 6(d) – (k).

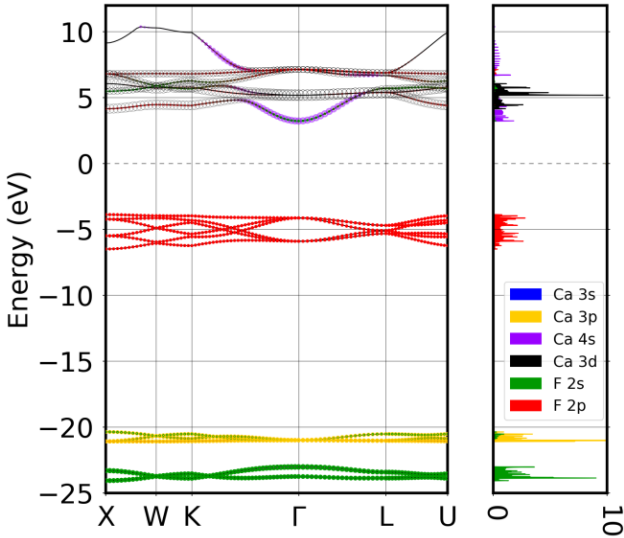

FIG 5. Calculated band structure of $CaF_2$ suggests the orbital shapes of core, valence, and conduction band levels involved during UXTAS experiment. The projected band structure and projected density of states were calculated by using DFT.

The BSE computations find that the peak of $X_1$ can be associated with a strongly localized Γ-point exciton (in agreement with previous work [9]), which is of Ca 4s character, primarily [Fig. 6(d)]. Towards the higher-energy edge of the $X_1$ peak, other localized CB edge excitons with lower oscillator strengths can be found such as the X-point exciton in Fig. 6(e), which is of Ca 3d character, primarily. In the region that corresponds to the location of $X_2$ in the experimental data, one identifies excitons that seem more delocalized across the lowest CB level. As a result, it would be expected to exhibit a mix of Ca 4s and Ca 3d orbital shape with specific contributions depending on the region within the $X_2$ feature. It is ascertained that the remaining features correspond to the broad $X_3$ feature in the experimental data. This absorption peak seems to comprise core excitons consisting of delocalized contributions across higher levels in the CB. In particular, the higher energy region of this range of excitons exhibits contributions from the levels that begin to exhibit some F 2p character in addition to the Ca 4s and Ca 3d character that most of the core excitons appear to show.

## IV. DISCUSSION

The experimental and computation results presented above present a cohesive explanation for the nature of the core exciton absorption peaks in the linear spectrum, their transfers undergone during the XUV pump - NIR probe time-resolved experiment, and their decoherence mechanisms. A discussion on each of these topics is considered below.

### A. Core excitonic peaks and transfer

The experimental linear absorption spectrum consists of three main features before the rise in absorption at 32 eV due to Ca 3p-to-CB and F 2s-to-CB excitation: $X_1$, $X_2$, and $X_3$. All core excitons comprising these absorption features result from excitation from the Ca 3p core levels. $X_1$ is concluded to result from excitation to a band edge Γ-point core exciton and exhibits Ca 4s orbital shape. $X_2$ which appears to be made up of various excitonic features, is concluded to consist of core excitons delocalized across the CB edge and is expected to exhibit mixed Ca 4s and 3d orbital shape. The broad $X_3$ absorption is also expected to consist of various core excitons, in this case delocalized across higher CB levels and is expected to exhibit mixed Ca 4s and 3d orbital shape. Interestingly the higher energy region of the $X_3$ feature also begins to exhibit F 2p orbital shape.

After the XUV pump, the 1.2-2.5 eV NIR probe enables core excitonic transfer between the Γ-point core exciton in $X_1$ and the higher energy features in $X_3$ that exhibit F 2p orbital shape. This is a weakly allowed transition given the mixed Ca 4s, Ca 3d, and F 2p orbital shape of the $X_3$ excitons, and its transfer is further hampered by the weaker NIR intensity towards the edges of its broad spectrum. This transfer is evidenced by the bleaching Z feature, as well as the sublinear energy shift of $X_1$ vs. NIR intensity. Transfer between the $X_2$ and $X_3$ excitons, and between the $X_1$ and $X_2$ could not be identified. This might result from the transitions being disallowed due to more similar orbital angular momenta or due to too small an energy spacing between these sets of excitons compared to the NIR probe energy. It is also possible that transfer involving the $X_2$ feature takes place but is not possible to distinguish due to the several overlapping excitons within this feature.

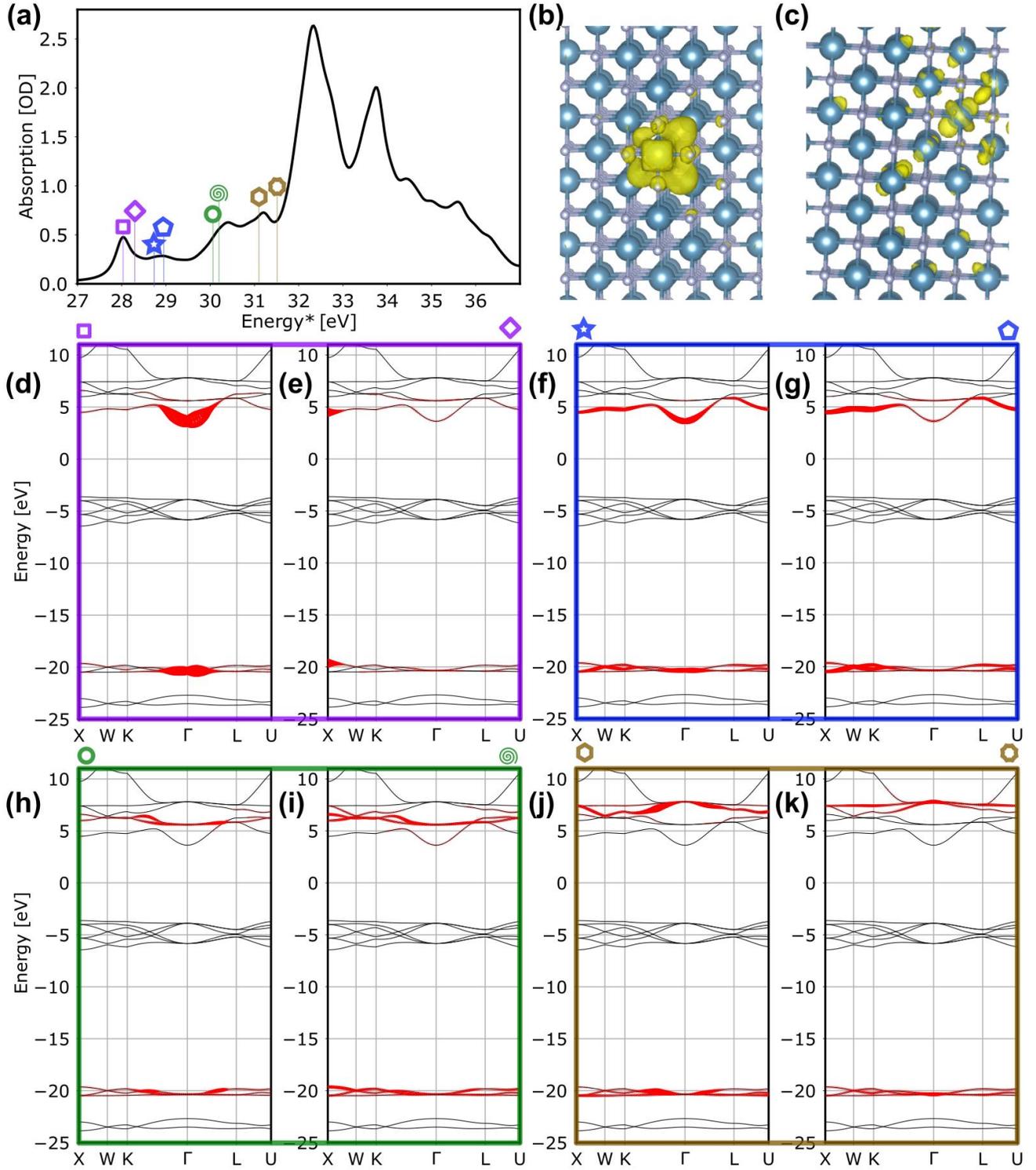

FIG 6. Exciton computational analysis confirms the modes involved in the excitonic features and transfer. a) Linear absorption spectrum with various representative excitons within each absorption peak. b, c) Excitonic wavefunctions for the purple square and diamond, respectively. d-k) Excitonic weights depicted in red plotted on the CaF$_2$ band structure for the highlighted features on the spectrum in sequential order, i.e. purple square, purple diamond, blue star, blue pentagon, green circle, green spiral, yellow hexagon and yellow eight-point star. Excitonic computational analysis performed with the Exciting computational package using the Bethe-Salpeter-equation (BSE).

We also note that it was unsuccessfully attempted to detect a transient ultrafast XUV signal due to the $CaF_2$ core excitons in a four-wave mixing experimental setup specialized for solid-state materials, similar to previous work [14]. This is attributed to weakly allowed transitions that will also lead to weak UXTAS signals, as is discussed earlier. A detailed account of the four-wave mixing experimental setup and results will be detailed in a future report.

### B. Phonon-mediated core exciton decoherence

Three relatively long-lived features are identified in the UXTAS data, labelled A, B, and C. The first is associated with the decoherence of the Γ-point exciton, whereas the latter two are associated with core excitons within $X_2$. All three excitonic features are directly excited by the initial XUV pump and then decohere over the course of a few fs. Their coherence lifetimes (spanning 3-7 fs for various excitons and intensities) are within the same range as those in MgO [12] and $SiO_2$ [11], where it was concluded that core excitons decohere as a result of electron-phonon coupling. These studies note that the phonon contribution to the dipolar phase decay can be described by (as detailed by Mahan [22]):

$$Im\left(\varphi_{ph}(t)\right) = \frac{M^2}{\omega_{ph}^2}[(2N+1)(1-cos\omega_{ph}t)],$$

which denotes Gaussian spectral lineshapes when $\omega_{ph}t \ll 1$. Other studies have cited Lorentzian lineshapes as indicative of alternative decay pathways [14,15]. As illustrated in Fig. 3, the spectral features of the $CaF_2$ core excitons are best fitted by Gaussian lineshapes. Furthermore, much like in MgO and $SiO_2$, the expected Auger Lorentzian linewidth for a Ca 3p core level excitation would be 10 meV [27], which would result in a lifetime far longer than what we observe in $CaF_2$. As a result, the observed exciton decoherence cannot be attributed to Auger decay.

Additionally, the measurement of the coherence lifetime at various NIR intensities shows a decrease in the lifetime at increased probe intensity, further supporting a phonon-mediated decay mechanism. As the NIR peak intensity is increased, so is the phonon population by electron-phonon coupling. It is estimated from the core exciton linear absorption linewidth (linewidth σ = 102 meV), after deconvolving the 10 meV Auger Lorentzian linewidth contribution, that approximately 6 phonons are generated per photon [28,29]. This was calculated by considering that the energy of the optical phonon involved would be between 31-55 meV [30]. Therefore, a large population of phonons will result from electron-phonon coupling upon excitation of the core-exciton with the XUV, which could be responsible for the primary decay component observed in $CaF_2$'s core excitons.

Thermal phonons are also a possibility as an alternative source of phonons that lead to the short coherence lifetimes observed here. The phonon population in $CaF_2$ will follow Bose-Einstein statistics and will increase as the temperature is increased [23]. Similarly, time-invariant features observed in UXTAS that are also present between pulses in 1 kHz-repetition-rate experiments have been previously attributed to heat features in the literature [31]. It is therefore plausible that the time-independent thermal phonon population is also increased due to the high repetition rate NIR beam, such that an additional thermal phonon population might also contribute to the core exciton decoherence. It is proposed that both electron-phonon coupling upon XUV excitation and a thermal phonon population that increases due to the high intensity NIR beam could contribute to core exciton decoherence. Further experiments need to be designed in order to distinguish each individual contribution.

## V. CONCLUSION

In summary, core excitons in $CaF_2$ at the $M_{2,3}$ edge were observed and studied by UXTAS. Various core excitons excited directly from the Ca 3p core levels were identified within the XUV linear absorption region between 27-37 eV. Excitonic transfer was ascertained between a bright Γ-point core exciton that exhibits Ca 4s orbital shape and a weakly visible core exciton that exhibits mixed F 2p, Ca 4s, and Ca 3d orbital shapes. The decay of the excitonic features is concluded to result from phonon-mediated decoherence, namely electron-phonon coupling. This study sought to understand the behavior of core excitons in $CaF_2$, as well as to illuminate the general behavior of core excitons in solid-state materials. Additionally, the joint approach involving experimental UXTAS and computational modelling consisting of projections onto the band structure is presented as a successful method to understand the rich behavior and interactions of core excitons in solids.

### ACKNOWLEDGEMENTS

This work was supported by the Air Force Office of Scientific Research (AFOSR), Grant FA9550-20-1-0334. R.Q.B. acknowledges support from Natural Sciences and Engineering Research Council of Canada (NSERC) Postdoctoral Fellowship.

### APPENDIX A: EXPERIMENTAL METHODS

A new apparatus was constructed to specifically measure attosecond XUV pump probe spectroscopy in solid-state materials and will be further detailed in a future report. The NIR and XUV pulses in the UXTAS experiment are produced by a Ti:sapphire laser (Coherent Astrella, 1 khz repetition rate, 7 mJ pulse energy, 35 fs pulse duration and 32 nm bandwidth). 3 mJ of the beam is focused with an f=2.5 m dielectric concave mirror into a 2.2 m long hollow core fiber (inner diameter 700 μm) filled with an Ar gas pressure gradient (approximately 5 Torr to 120 mTorr) to generate a supercontinuum spanning 500–1000 nm wavelength with self-phase modulation. Eight pairs of double angled broadband chirped mirrors (Ultrafast Innovations, PC1332), fused silica wedge pairs, and potassium dihydrogen phosphate (KDP) wedge pairs are optimized to compensate accumulated dispersion and temporally compress the pulses to ~5 fs durations. The beam is then split by a 90:10 broadband beamsplitter. The former is subsequently focused into a gas cell with a 200 μm hole with Kr gas at ~15 Torr in a vacuum chamber held at ~$10^{-7}$ Torr to generate broadband XUV pulses via HHG. A 100 nm Al foil filter (Lebow Company) is then used to filter out the NIR and transmit the XUV pulse, which is focused onto the sample with a gold-coated toroidal mirror. The remaining NIR beam after the beamsplitter is used as the probe beam. The NIR probe is delayed with respect to the XUV beam with a piezo stage (Physik Instrumente, P-620.1CD). The NIR probe is then focused into the vacuum chamber system with an f=1 m silver concave mirror. The beam is incident onto the sample off-axis from the XUV beam by a small angle ($\theta$=23 mrad) in order to focus both beams onto the sample accurately without clipping. The pulse energy of the NIR beam is adjusted with an iris and monitored with a power meter between 6-12 µJ. The NIR beam diameter is characterized using a beam profiler (DataRay WinCamD). Temporal and spatial overlap is determined by placing an Ar gas cell in the interaction region instead of the sample. The NIR pulse width is estimated to be under 5 fs with a dispersive characterization of ultrafast pulses (D scan, Sphere Ultrafast Photonics) [Fig. S1] and confirmed with the rise time of the Ar $3s3p^6np$ autoionization signal. The spectral energies are also calibrated with the autoionization lines of Ar $3s3p^6np$ [32]. Another 100 nm Al foil is then used to filter out any remaining NIR, meanwhile the XUV beam is spectrally dispersed by a gold-coated flat-field grating (01-0639, Hitachi) and measured with an XUV CCD camera (Pixis XO 400-B, Princeton Instruments).

The samples measured are 25 nm polycrystalline $CaF_2$ thin films (Lebow Company) evaporated onto 30 nm thick $Si_3N_4$ membranes (Norcada NX5050X). The samples are mounted on a xy-axis translation stage (Physik Instrumente). The illuminated area is refreshed every 5 minutes using vertical and horizontal sample translation to avoid artifacts from sample heating and damage.

**APPENDIX B: COMPUTATIONAL METHODS**

The projected band structure of $CaF_2$ is calculated using the plane-wave DFT software package Quantum Espresso. We use Perdew-Burke-Ernzerhof (PBE) projector augmented wave (KJPAW) functionals, and a *k* grid of 8x8x8. The excited state band structure projections are computed using exciting, a full-potential all-electron package implementing linearized augmented planewave methods. PBEsol functionals, and a 8x8x8 *k* grid are used. Excited state properties are calculated with the Bethe-Salpeter-equation and visualized with exciting. For the Bethe-Salpeter-equation calculations, 6 core levels (Ca 3s, Ca 3p, and F 2s), all 6 VB levels, and 6 CB levels are used to calculate all excited states within the 27-37 eV absorption region (after scissor correction). A scissor correction is used to match the DFT-computed absorption spectrum with the experimental spectrum to make up for DFT's underestimation of the bandgap.